%% file: main.tex
\documentclass[conference]{IEEEtran}
\IEEEoverridecommandlockouts
\usepackage{cite}
\usepackage{amsmath,amssymb,amsfonts}
\usepackage{algorithmic}
\usepackage{graphicx}
\usepackage{textcomp}
\usepackage{xcolor}
\usepackage{subcaption}
\usepackage{multirow}
\usepackage{pifont}
\usepackage{hyperref}
\newcommand{\cmark}{\ding{51}}%
\newcommand{\xmark}{\ding{55}}%
\def\BibTeX{{\rm B\kern-.05em{\sc i\kern-.025em b}\kern-.08em
		T\kern-.1667em\lower.7ex\hbox{E}\kern-.125emX}}
\begin{document}
	
\title{CWcollab: A Context-Aware Web-Based Collaborative Multimedia System}

\author{\IEEEauthorblockN{Chunxu Tang* \thanks{* The work was done when the authors were at Syracuse University, Syracuse, USA.}}
	\IEEEauthorblockA{
		\textit{Twitter, Inc}\\
		San Francisco, USA \\
		chunxut@twitter.com}
	\and
	\IEEEauthorblockN{Beinan Wang*}
	\IEEEauthorblockA{
		\textit{Twitter, Inc}\\
		San Francisco, USA \\
		beinanw@twitter.com}
	\and
	\IEEEauthorblockN{C.Y. Roger Chen}
	\IEEEauthorblockA{\textit{Department of EECS} \\
		\textit{Syracuse University}\\
		Syracuse, USA \\
		crchen@syr.edu}
	\and
	\IEEEauthorblockN{Huijun Wu}
	\IEEEauthorblockA{\textit{Department of CIDSE} \\
		\textit{Arizona State University}\\
		Tempe, USA \\
		huijun.wu@asu.edu}
}

\maketitle

\begin{abstract}

Remote collaboration tools for conferencing and presentation are gaining significant popularity during the COVID-19 pandemic period. Most prior work has issues, such as a) limited support for media types, b) lack of interactivity, for example, an efficient replay mechanism, c) large bandwidth consumption for screen sharing tools. In this paper, we propose a general-purpose multimedia collaboration platform-CWcollab. It supports collaboration on general multimedia by using simple messages to represent media controls with an object-prioritized synchronization approach. Thus, CWcollab can not only support fine-grained accurate collaboration, but also rich functionalities such as replay of these collaboration events. The evaluation shows hundreds of kilobytes can be enough to store the events in a collaboration session for accurate replays, compared with hundreds of megabytes of Google Hangouts.

\end{abstract}


\input{text/introduction}
\input{text/related-work}

\input{text/architecture}

\input{text/evaluation}

\input{text/conclusion}

\bibliographystyle{IEEEtran}
\bibliography{library}
	
\end{document}

%% file: text/introduction.tex
\section{Introduction}\label{sec.intro}

Real-Time Collaboration (RTC) has a long history in both research and engineering domains. It is gaining significant popularity during the COVID-19 pandemic period. For geographically dispersed teams, it is usually essential for teammates to have a remote collaboration tool for conferencing and presentation purposes. Additionally, considering the distributed and rapidly increasing volume of data and expeditious development of modern web browsers, it is necessary to provide an organized web-based group-aware platform to support collaboration among a large number of users. In RTC platforms, not only are the artifacts of multimedia shared but also the controls on the artifacts are broadcast and synchronized.

Some of the pioneering work, such as GroupSketch~\cite{greenberg1990group}, VideoWhiteboard~\cite{tang1991videowhiteboard}, and Liveboard~\cite{elrod1992liveboard}, implemented collaborative multimedia systems by capturing users' drawings and projecting them to remote screens. Based on the same technology, various screen sharing products, such as Zoom, Cisco WebEx, and Google Hangouts/Meet, were developed and are now widely used. Recently, with the rapid development of modern web technologies, lots of web-based collaboration tools, including Collabode~\cite{goldman2011real}, RichReview++~\cite{yoon2016richreview++}, and Tele-Board~\cite{wenzel2016full}, were developed.

To our knowledge, even though various web-based groupware tools have been developed for versatile purposes such as whiteboard drawing and document editing, no systems can integrate these functionalities to suit general-purpose multimedia collaboration. Kim et al.~\cite{kim2015interactive} made a valuable contribution in this direction that they proposed an MVC architecture for ubiquitous collaboration. Their tool still only handled static media like whiteboard drawings and images, without effectively working on other types of media with dynamic contents such as videos and web pages.

Furthermore, for current popular screen sharing products, in a specific session, both the contents of the media and manipulations on the media are transmitted by capturing the display continuously, leading to large consumption of network bandwidth. This poses challenges for large geographically dispersed teams with the unstable network quality. By contrast, we propose to split the contents of a presentation into static media resources and dynamic actions. The static media resources, for example, a video or a PDF document, can be transmitted to attendees beforehand; and the dynamic events occurring in a session such as muting a video are broadcast and synchronized on the fly. As a result, a collaboration session is organized as the combination of static materials and dynamic events encapsulated in an event-driven stream of messages. With these messages, we also implemented the precise recording and replay of collaboration events, differentiating our work from traditional collaboration platforms. 

To cover the research gaps in prior work mentioned above, we propose a context-aware web-based collaborative multimedia system-CWcollab. Specifically, our contributions are:

\begin{enumerate}
	\item \textbf{A general-purpose collaborative multimedia system. }
	\begin{itemize}
		\item Support for general multimedia. Not only static media (PDF documents, images, etc.) but also dynamic media (videos, web pages, etc.), is supported in CWcollab. This notable feature differentiates our work from prior work, as illustrated in Table~\ref{tab:comparison-media}.
		
		\item Support for general events. All actions in a session are captured as events, sent and handled on the fly. This also indicates the system is open to other possible extensions in a plugin pattern. A developer can add supports for various events of collaboration on different kinds of media following our uniform interface, demonstrated in Section~\ref{sec.arch.design}.
		
		\item Support for general environment. Our system is totally web-based. Users can access the system from various platforms, including desktops, tablets, and mobile devices, as long as web browsers are supported. This significantly eliminates the complexity of setup on different platforms.
	\end{itemize}

	\item \textbf{An object-prioritized context-aware approach to capture and replay media actions for rich functionalities with low network bandwidth.} 
	\begin{itemize}
		\item Object-prioritized media controls. To our knowledge, most web-based collaboration tools are position-based or proportion-based, which implies that media controls are synchronized through absolute or relative positions. However, many websites have applied the responsive web design, where the user interface is automatically adjusted on various devices with various screen sizes. This poses challenges for capturing and replaying events with the traditional position-based approach. By contrast, in CWcollab, media controls are related to media objects. We propose an object-prioritized hybrid synchronization approach, representing each action in simple messages, discussed in Section~\ref{sec.arch.capturer}. 
		
		\item Rich functionalities. As each media control is represented by a simple message, CWcollab also supports rich interactive functionalities for collaboration and presentation such as material preparation, real-time synchronization, and precise replay of a session.
		
		\item Low network bandwidth. The usage of an object-prioritized approach also implies that CWcollab has a very low bandwidth usage, compared with current video conferencing products such as Zoom and Google Hangouts. In our study, hundreds of kilobytes can be enough to store the events in a session, compared with hundreds of megabytes used in screen sharing tools. 
	\end{itemize}

\end{enumerate}

The remainder of this paper introduces the related work in Section~\ref{sec.related-work}, discusses the architectural design and implementation in Section~\ref{sec.arch}, and evaluates our platform including a comparison with Google Hangouts in Section~\ref{sec.eval}. Section~\ref{sec.conclusion} concludes the paper.

%% file: text/related-work.tex
\section{Related Work}\label{sec.related-work}

RTC is usually considered as a subdomain of Computer Supported Cooperative Work. Previously, people achieved basic remote collaboration by means of video/audio calls. While it is obvious that this form of communication can only allow very preliminary cooperation. Afterward, researchers developed more complicated collaborative applications. Some of the pioneering work includes GroupSketch~\cite{greenberg1990group}, VideoWhiteboard~\cite{tang1991videowhiteboard}, and Liveboard~\cite{elrod1992liveboard}. In these preliminary systems, users and their drawings are captured by cameras, transmitted and projected to remote screens. An issue of these systems is that only screenshots are transmitted, thus they lack the flexibility to provide interaction among people geographically dispersed. Subsequently, researchers developed more elaborate native desktop collaborative applications. For instance, Booth et al.~\cite{booth2002mighty} proposed a ``mighty mouse" multi-screen collaboration tool, which provides a smooth mouse movement cross-platform, via VNC protocol. A crucial issue of this kind of systems is that they usually require complicated setups on different platforms because they need to have distinct implementations on these platforms.

Recently, with the significant enhancement of modern web browsers, especially with the advent of technologies brought from HTML5 standard, such as WebSocket, researchers began to consider web browsers as the backbone for RTC applications. Bentley et al.~\cite{bentley1997designing, bentley1997world} gave pioneering work in this direction. They produced a basic shared workspace system across web browsers. Mogan et al.~\cite{mogan2010impact} performed a comparison study on three group-aware tools: desktop-based Java tool XCHIPS~\cite{wang2009cooperative}, browser-based Adobe Flash application ThinkTank, and browser-based AJAX system PowerMeeting. They claimed that the last one has advantages in features, user interface, usefulness, etc. over the others. Schmid et al.~\cite{schmid2014real} employed an XML format to develop a web-based interactive collaborative environment. Now, Google Docs and Etherpad are both rather mature products which enable real-time document editing collaboration among multiple users. Fetter et al.~\cite{fetter2014lightweight} created a collaborative web browsing tool distilling the concept of lightweight interference, transitions, and adoptions. Binda et al.~\cite{binda2018phamilyhealth} developed a photo-sharing system to stay aware of family members' health. Lee et al.~\cite{lee2018exploring} developed a UI design tool to explore RTC in crowd-powered systems.

\begin{table*}[]
	\centering
	\caption{Comparison of multimedia collaboration system features.}
	\label{tab:comparison-media}
	\begin{tabular}{|l|c|c|c|c|c|c|}
		\hline
		\textbf{System} & \begin{tabular}[c]{@{}c@{}} \textbf{Document editing}\end{tabular} & \begin{tabular}[c]{@{}c@{}}\textbf{Web browsing}\end{tabular} & \textbf{Whiteboard} & \begin{tabular}[c]{@{}c@{}}\textbf{Image annotation}\end{tabular} & \begin{tabular}[c]{@{}c@{}}\textbf{Video watching}\end{tabular} &
		\begin{tabular}[c]{@{}c@{}}\textbf{Video annotation}\end{tabular} \\ \hline
		Google Docs                                & \cmark                                                      & \xmark                                                  & \xmark     & \xmark                                                      & \xmark  & \xmark                                                  \\ \hline
		Google Hangouts & \xmark & \xmark & \xmark & \xmark & \cmark & \xmark \\ \hline
		SpreadVector \cite{fetter2014lightweight} & \xmark                                                      & \cmark                                                  & \xmark     & \xmark                                                      & \xmark     & \xmark                                               \\ \hline
		Collaboard \cite{kunz2010collaboard}      & \xmark                                                      & \xmark                                                  & \cmark     & \xmark                                                      & \xmark     & \xmark                                               \\ \hline
		Kim et al. \cite{kim2015interactive}      & \xmark                                                      & \xmark                                                  & \cmark     & \cmark                                                      & \xmark      & \xmark                                              \\ \hline
		CWcollab                                  & \xmark*                                                           & \cmark                                                  & \cmark     & \cmark                                                      & \cmark        & \cmark                                            \\ \hline
		\multicolumn{7}{l}{*: Although CWcollab does not support fully-functional collaborative document editing, it does support collaborative annotation and} \\
		\multicolumn{7}{l}{manipulations on PDF documents.}
	\end{tabular}
\end{table*}

In our study of four prior collaboration systems, they only support very limited types of media, as demonstrated in Table~\ref{tab:comparison-media}. Here, Google Docs supports document editing, Google Hangouts supports video watching, SpreadVector~\cite{fetter2014lightweight} is for collaborative web browsing, and Collaboard~\cite{kunz2010collaboard} has a whiteboard functionality. The system described in~\cite{kim2015interactive} works for more types of media, even though it still only supports whiteboard and image editing. By contrast, CWcollab supports all types of multimedia listed in the table, especially for presentation purposes. This notable feature differentiates our work from prior work.

%% file: text/architecture.tex
\section{Architectural Design and Implementation}\label{sec.arch}

\subsection{High Level Design}\label{sec.arch.design}

\begin{figure}[!htb]
	\centering
	\includegraphics[scale=0.44]{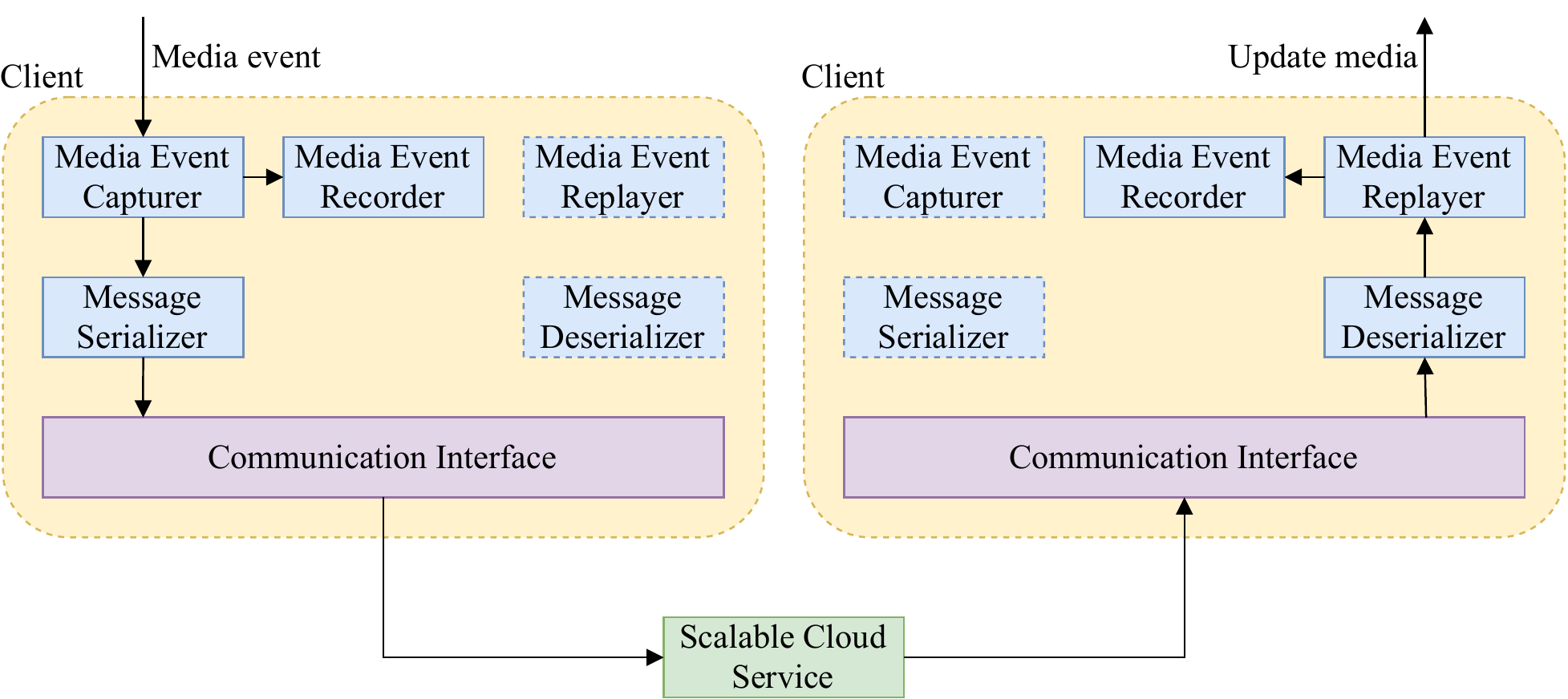}
	\caption{The implementation of collaboration based on a uniform interface.}
	\label{fig:uniform-interface}
\end{figure}

A crucial feature distinguishing our design from other collaboration systems is the general-purpose distributed collaboration with a uniform interface. As events are transmitted in a standardized message format, this design significantly relieves the difficulty of supporting a new type of media. 
Figure~\ref{fig:uniform-interface} shows the structure of collaboration based on a uniform interface.

\textbf{Sender flow.} For the client who initiates a media event for collaboration, there is a \textit{media event capturer} module to monitor and capture the event. A \textit{media state recorder} will be invoked when necessary to record the current media state for possible resynchronization. For example, the state of a video may include the play/pause state, volume, progress, playback rate, related annotations, etc. At the same time, the event is sent to a \textit{message serializer}, which encapsulates the event into a standardized message and sends it to the backend service. 

\textbf{Receiver flow.} After the message is routed to one target client via the backend service, a \textit{message deserializer} takes charge of unwrapping the message and sending information to the \textit{media event replayer} to update the current media state. Simultaneously, the change of media state may also be recorded in the \textit{media state recorder} module.

With the uniform interface, if a developer would like to add collaboration mechanisms for a new type of media, he/she just needs to follow the structure by adding functions in the \textit{media event capturer}, \textit{media state recorder}, and \textit{media event replayer}. \textit{Message serializer} and \textit{message deserializer} should be left intact if the format of transmitted messages is not changed. The feature of extendability significantly relieves the burden of implementing new functionalities for collaboration.

\subsection{Media Event Capturer}\label{sec.arch.capturer}

\begin{figure}[htb]
	\centering
	\includegraphics[scale=0.33]{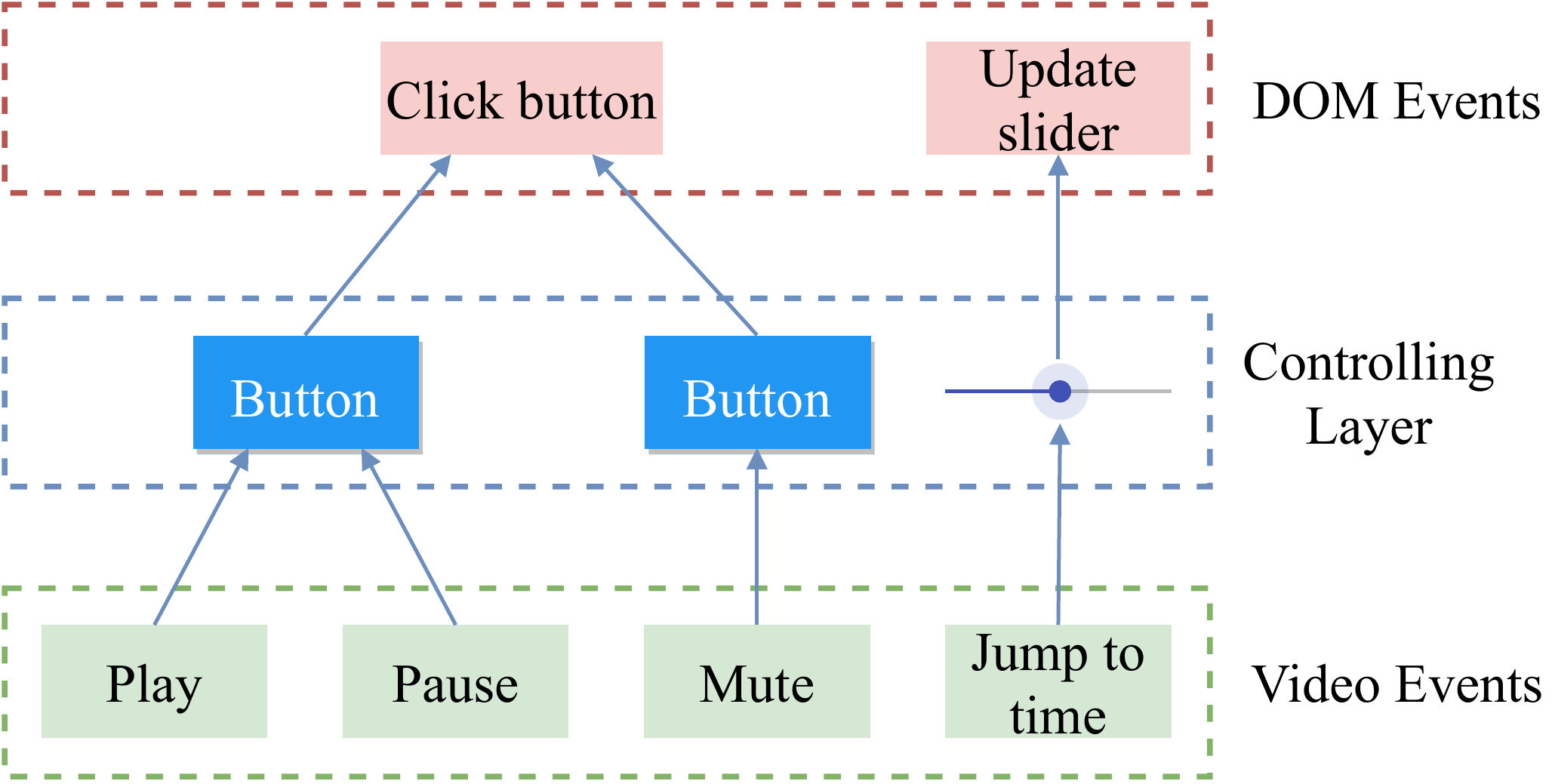}
	\caption{The structure of capturing some semantic video events in DOM events.}
	\label{fig:video-event-capturing}
\end{figure}

The main functionality of the \textit{media event capturer} component is to capture an abstract media event via DOM events. In our design, by using event propagation, especially event bubbling, which is now supported in all of the modern browsers, we create another layer on top of original media objects to capture these events. We leverage an object-prioritized hybrid approach to capture media events. Specifically, it consists of:

\begin{itemize}
	\item Object-based. Events are captured as DOM events on the media object or a UI component. This provides more flexibility than the traditional position-based approach, as no matter where a media event fires, we only trace the object effected, isolated from the influence of positions. This type of capture scenario is much more often seen than others. Figure~\ref{fig:video-event-capturing} illustrates how to capture object-based video events. In the figure, every control is triggered through a UI component. For example, when a user plays/pauses a video, he/she clicks the video control button, which is eventually captured by the \textit{media event capturer} as a \textit{click button} DOM event. similarly, when a user seeks to a timestamp in the video player, he/she clicks the progress bar, and the \textit{media event capturer} captures this event as an \textit{update slider} DOM event.
	
	\item Proportion-based. Events are captured as DOM events related to the positions proportionally to the display. Although object-based capture handles most scenarios, some media events may not be object-based, especially those related to mouse events. For example, we use the relative position of an image to the screen to synchronize an \textit{image moving} event.
	
	\item Value-based. Events are captured as DOM events related to value changes. For example, when scrolling a mouse to zoom in/out an image, the event can be captured via a \textit{wheel}, \textit{mousewheel}, or \textit{DOMMouseScroll} event. This type of event is fired with a value to represent how much the mouse has scrolled. 
\end{itemize}

\subsection{Media Event Recorder and Replayer}

The \textit{media event recorder} is used to track the current state of the media. The state of a media block depends on the type of media. A state is a set of key-value pairs. For example, a state of a video contains the video source, current timestamp, muted or not, volume, playback rate, annotations, etc. The key-value pairs can be compressed into messages for further transmission.

\begin{figure}[htb]
	\centering
	\includegraphics[scale=0.15]{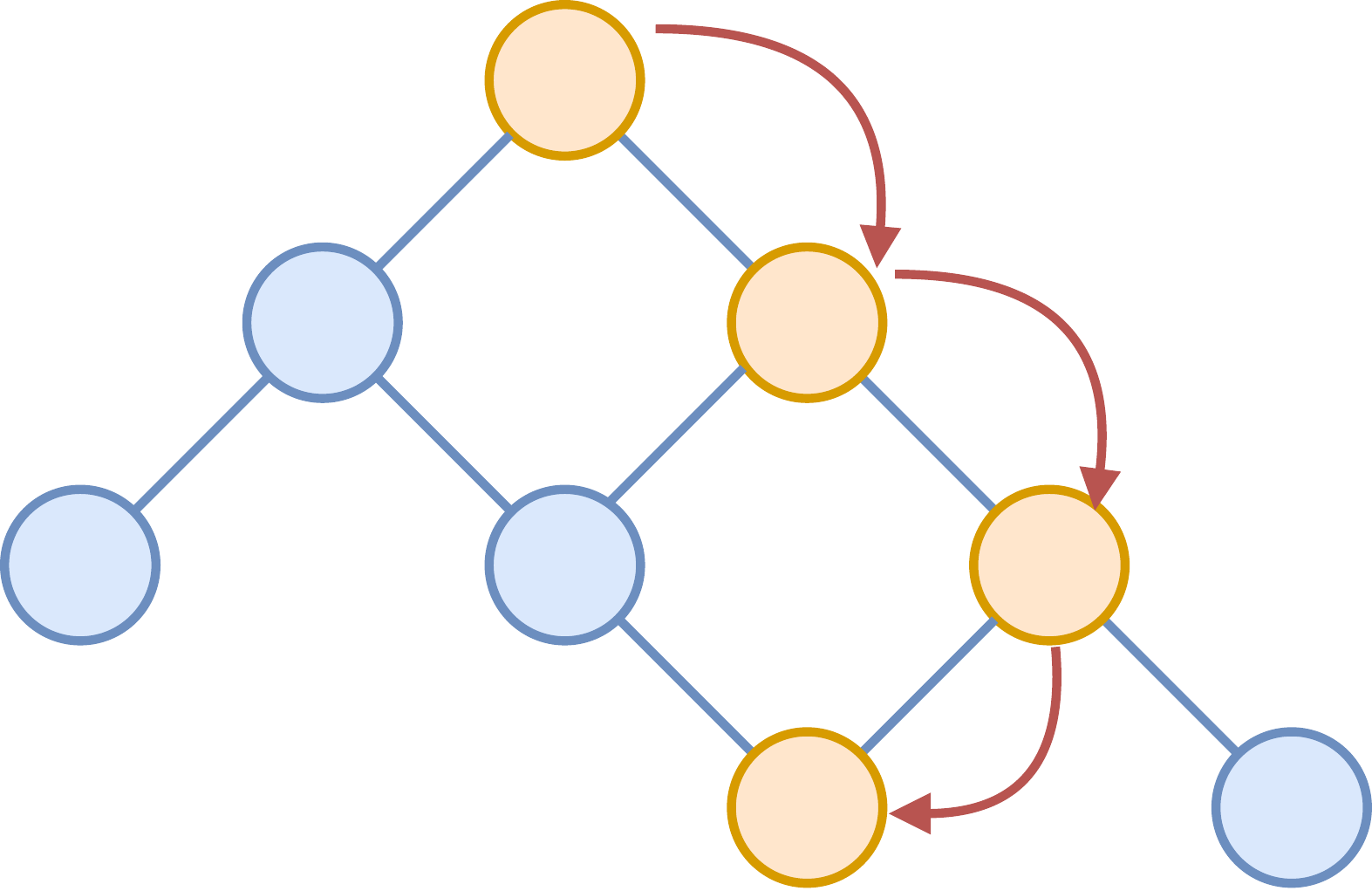}
	\caption{Propagation of event handling in a handler tree.}
	\label{fig:event-handler-propogate}
\end{figure}

The main functionality of a \textit{media event replayer} is to replay events attached with complete media event information. To achieve that efficiently, we design a hierarchical tree structure-\textit{Handler Tree}-to handle messages, as shown in Figure~\ref{fig:event-handler-propogate}.

\begin{figure}[htb]
	\centering
	\includegraphics[scale=0.376]{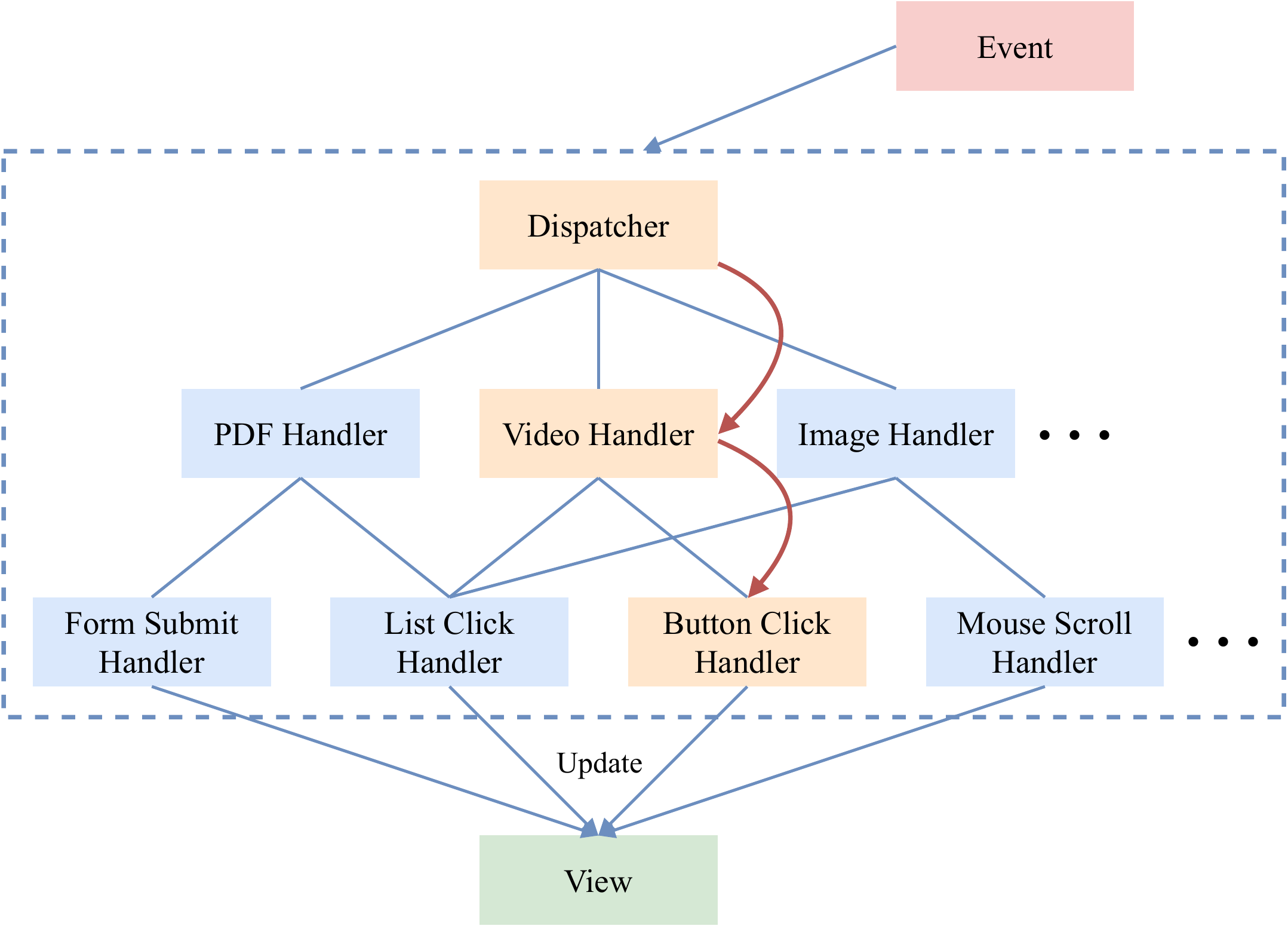}
	\caption{Propagation of event handling of a video \textit{play/pause} event in CWcollab.}
	\label{fig:event-handler}
\end{figure}

An event is propagated in the handler tree via parent-child chains until arriving at a leaf node. During the propagation, the received message is parsed gradually, and finally, in a leaf handler, the message is totally consumed to update the UI if necessary. An example of handling a video \textit{play/pause} event in CWcollab is demonstrated in Figure~\ref{fig:event-handler}.

When an event is received by an audience, it is first sent to the root node, also acting as a dispatcher in the handler tree. The next level in the handler tree contains various handlers for corresponding types of media. The message is sent to the appropriate media handler at this level. Then, a specific action handler is invoked to fire the target DOM event. For the example in Figure~\ref{fig:event-handler}, the message is first handled by the root handler and dispatched to the \textit{video handler} according to its media type. Because the event type is \textit{button click}, the message is next sent to the \textit{button click handler} for further process. Then the handler reads the target id and triggers the click event on the element with this identifier. The handler also loads the value of current time in the data field to update the current progress of the video to that timestamp.  Considering that our system is real-time, the tree is flat to reduce the overhead. 

With the help of the \textit{media event replayer}, the events occurring in a collaboration session can be replayed sequentially according to their timestamps. The events occurring in a session are pushed into a queue structure. These events are scheduled to be popped from the queue and sent to the handler tree for replay.

\subsection{Event Messages and Message Serializer/Deserializer}

In CWcollab, collaboration events are captured and represented by a standardized format of strings, and the strings are transferred among collaboration subjects. There are two different types of messages: \textbf{media events} and \textbf{control events}. For a media event message, it represents an action related to a specific media block, such as scrolling a web page. Table~\ref{tab:media-event-msg} demonstrates the possible fields in a media event message. By contrast, a control event message represents an action involving controls in a collaboration session, such as resynchronizing the media state. Table~\ref{tab:control-event-msg} shows the fields for a control event message.

\begin{table}[htb]
	\caption{Summary of elements in a media event message with an example for zooming out an image.}
	\label{tab:media-event-msg}
	\centering
	\begin{tabular}{|l|c|c|}
		\hline
		\textbf{Element}               & \textbf{Field}       & \textbf{Example}             \\ \hline
		media type identifier & media-type  & image               \\ \hline
		media identifier      & media-id    & image-block         \\ \hline
		event type identifier & event-type  & mouse-scroll        \\ \hline
		sequence identifier   & seq-id      & 8                   \\ \hline
		timestamp             & timestamp   & 5000                \\ \hline
		semantic              & description & zoom out an image    \\ \hline
		optional data         & data        & delta: -1.5 \\ \hline
	\end{tabular}
\end{table}

\begin{table}[htb]
	\caption{Summary of elements in a control event message with an example for resynchronizing a media block.}
	\label{tab:control-event-msg}
	\centering
	\begin{tabular}{|l|c|c|}
		\hline
		\textbf{Element}                        & \textbf{Field}                 & \textbf{Example}                                        \\ \hline
		control type identifier        & control-type          & resync                                         \\ \hline
		sequence identifier            & seq-id                & 5                                              \\ \hline
		timestamp                      & timestamp             & 10000                                          \\ \hline
		semantic                       & description           & Resync a media block                           \\ \hline
		\multirow{2}{*}{optional data} & \multirow{2}{*}{data} & media-id: video-block                          \\ \cline{3-3} 
		&                       & \multicolumn{1}{l|}{media-state: state string} \\ \hline
	\end{tabular}
\end{table}

The responsibilities of a \textit{message serializer} and \textit{message deserializer} are to wrap a media message into a standardized format of string and to read the message from such a string, respectively. To maintain consistency, the two components share the same message format.

%% file: text/evaluation.tex
\section{Evaluations}\label{sec.eval}

\subsection{Web Application}

We created a web application\footnote{ Demonstration video: \url{https://youtu.be/X-IWFSBvfSc}.} based on CWcollab to provide a collaborative presentation service, including account management, materials preparation, presentation synchronization, and replay. The backend service is implemented in Node.js, which is an event-driven, non-blocking, and cross-platform JavaScript run-time environment. We use MongoDB, a document-oriented NoSQL database, to store presentation data. We also use Redis, which is an in-memory key-value database, as a message broker.

The participants of a collaboration session can download media materials beforehand or the presenter can dynamically add materials in a session. In the application, manipulations on the material in the presentation panel are synchronized to other users' panels. 

\subsection{Comparison with Screen Sharing Tools}

\newcommand{\evalfigwidth}{0.23}
\newcommand{\lastfigwidth}{0.32}

\begin{figure*}[thb]
	\centering
	\begin{subfigure}[t]{\evalfigwidth\textwidth}
		\centering
		\includegraphics[width=0.8\linewidth]{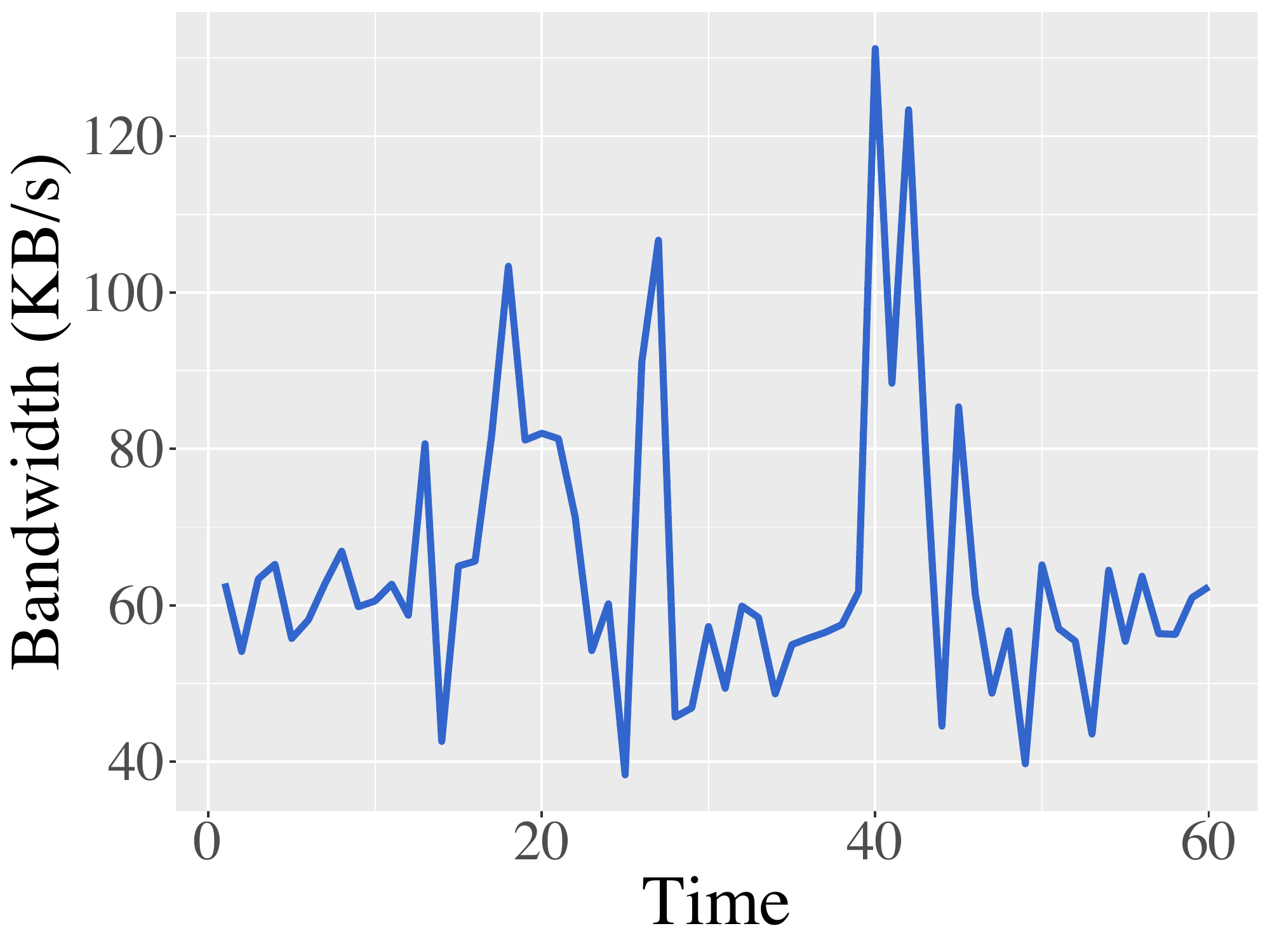}
		\caption{Google Hangouts.}
		\label{fig:video-google-preload}
	\end{subfigure}
	~
	\begin{subfigure}[t]{\evalfigwidth\textwidth}
		\centering
		\includegraphics[width=0.8\linewidth]{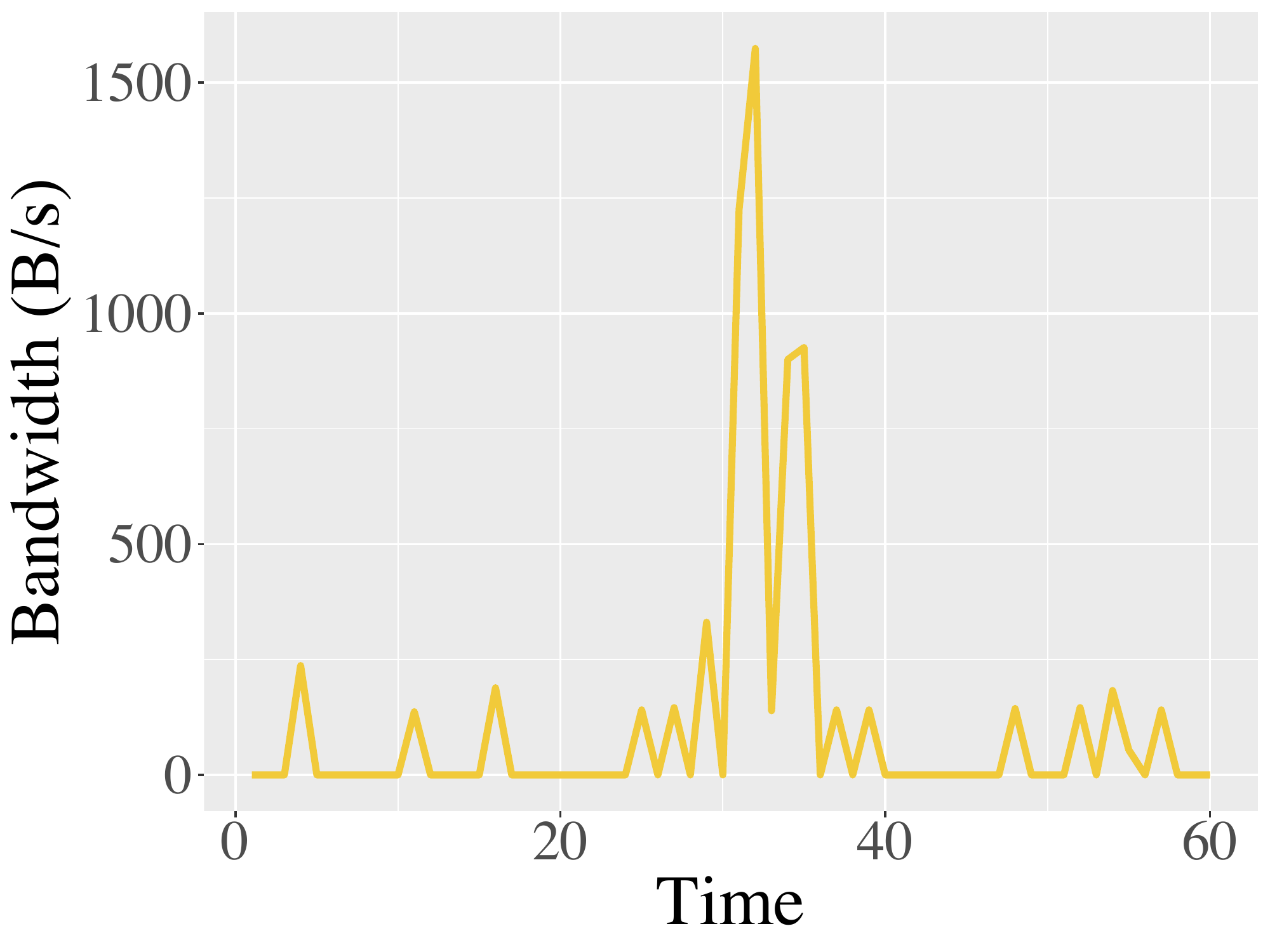}
		\caption{CWcollab.}
		\label{fig:video-our-preload}
	\end{subfigure}
	~
	\begin{subfigure}[t]{\evalfigwidth\textwidth}
		\centering
		\includegraphics[width=0.8\linewidth]{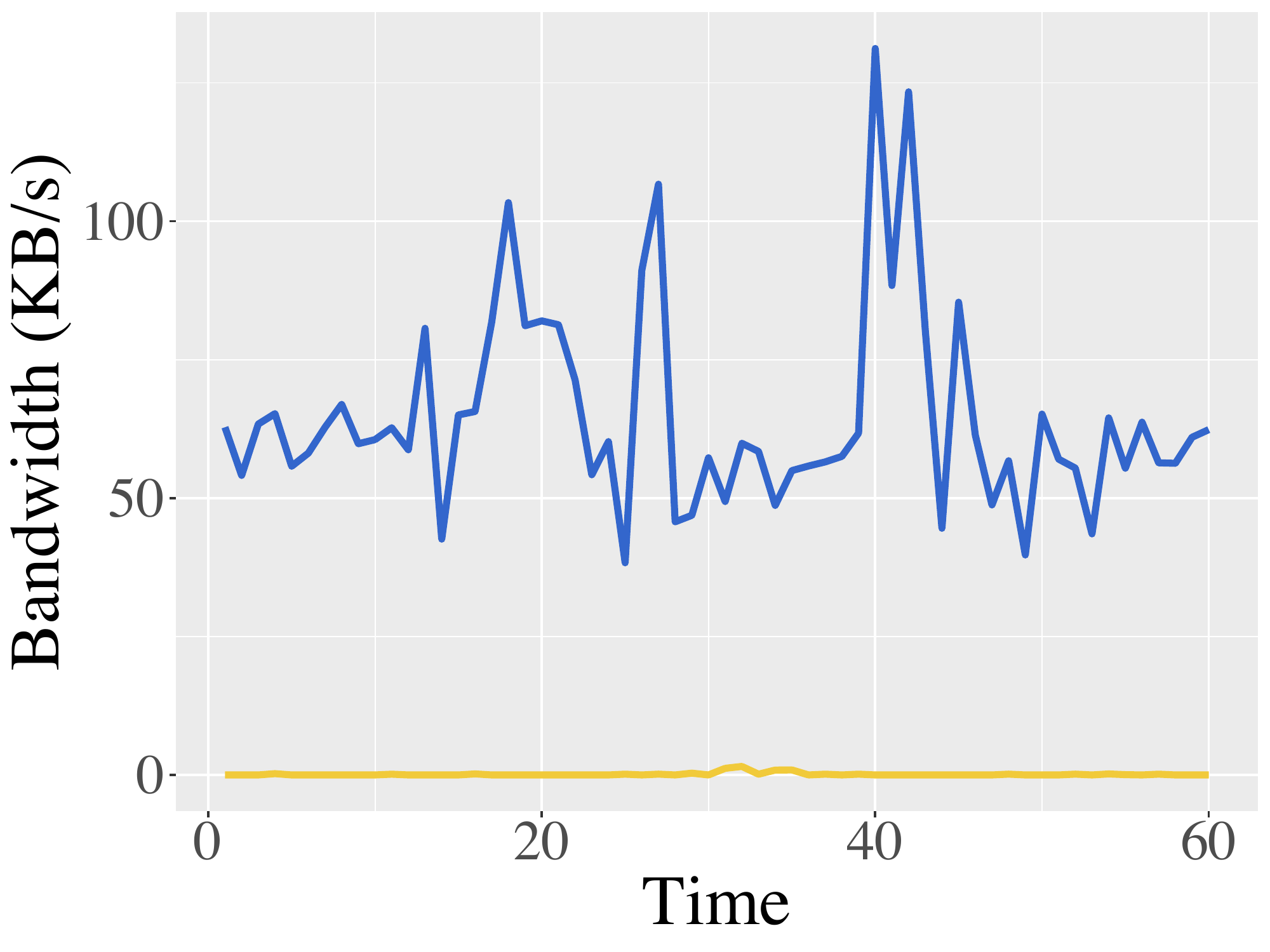}
		\caption{Comparison.}
		\label{fig:video-all}
	\end{subfigure}
	\caption{Comparison of network usages when presenting a \textbf{video} with operations such as playing, pausing, and free drawing using Google Hangouts and CWcollab.}
	\label{fig:video-eval-perf} 
\end{figure*}

\begin{figure*}[thb]
	\centering
	\begin{subfigure}[t]{\evalfigwidth\textwidth}
		\centering
		\includegraphics[width=0.8\linewidth]{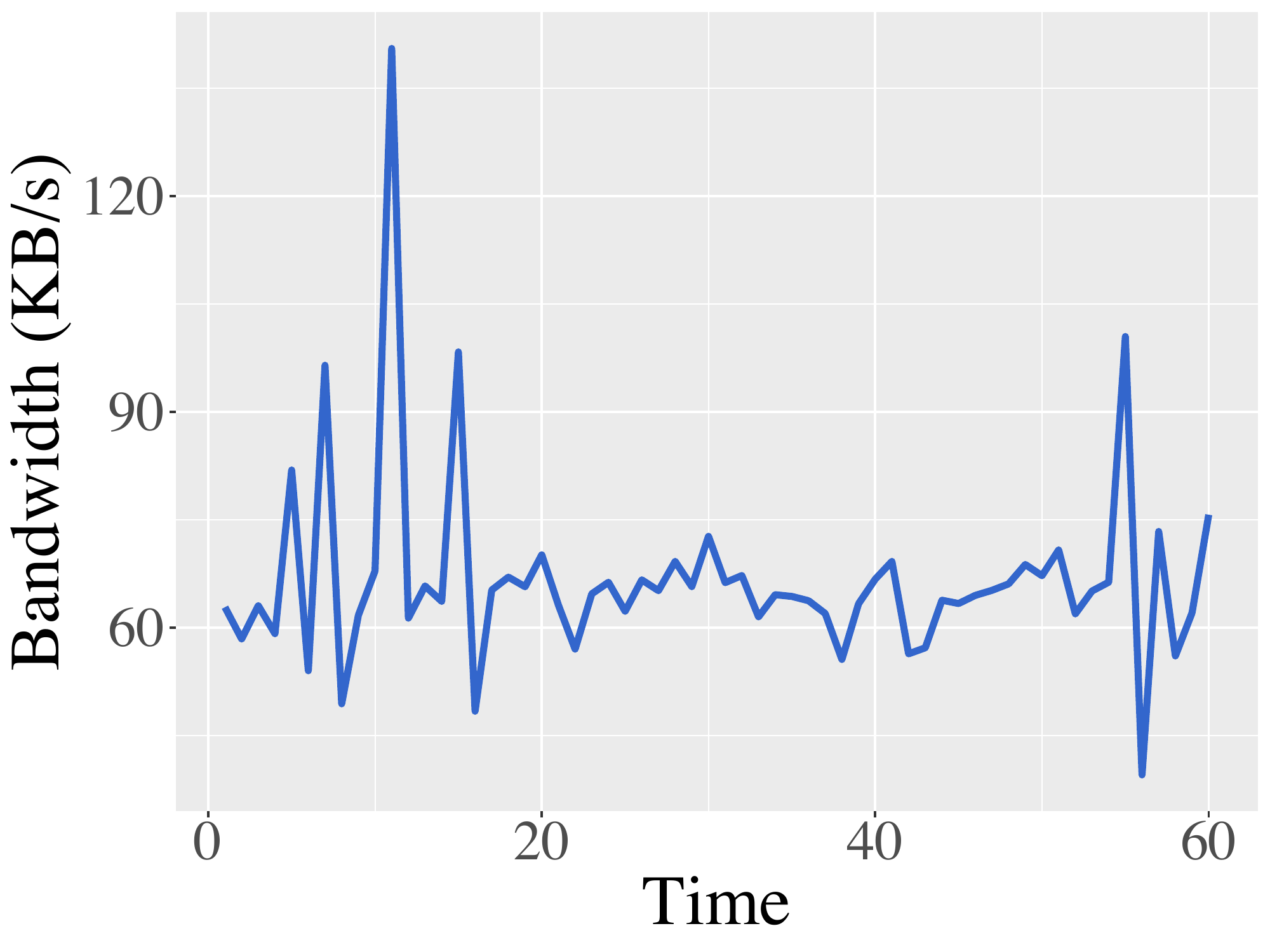}
		\caption{Google Hangouts.}
		\label{fig:pdf-google-preload}
	\end{subfigure}
	~
	\begin{subfigure}[t]{\evalfigwidth\textwidth}
		\centering
		\includegraphics[width=0.8\linewidth]{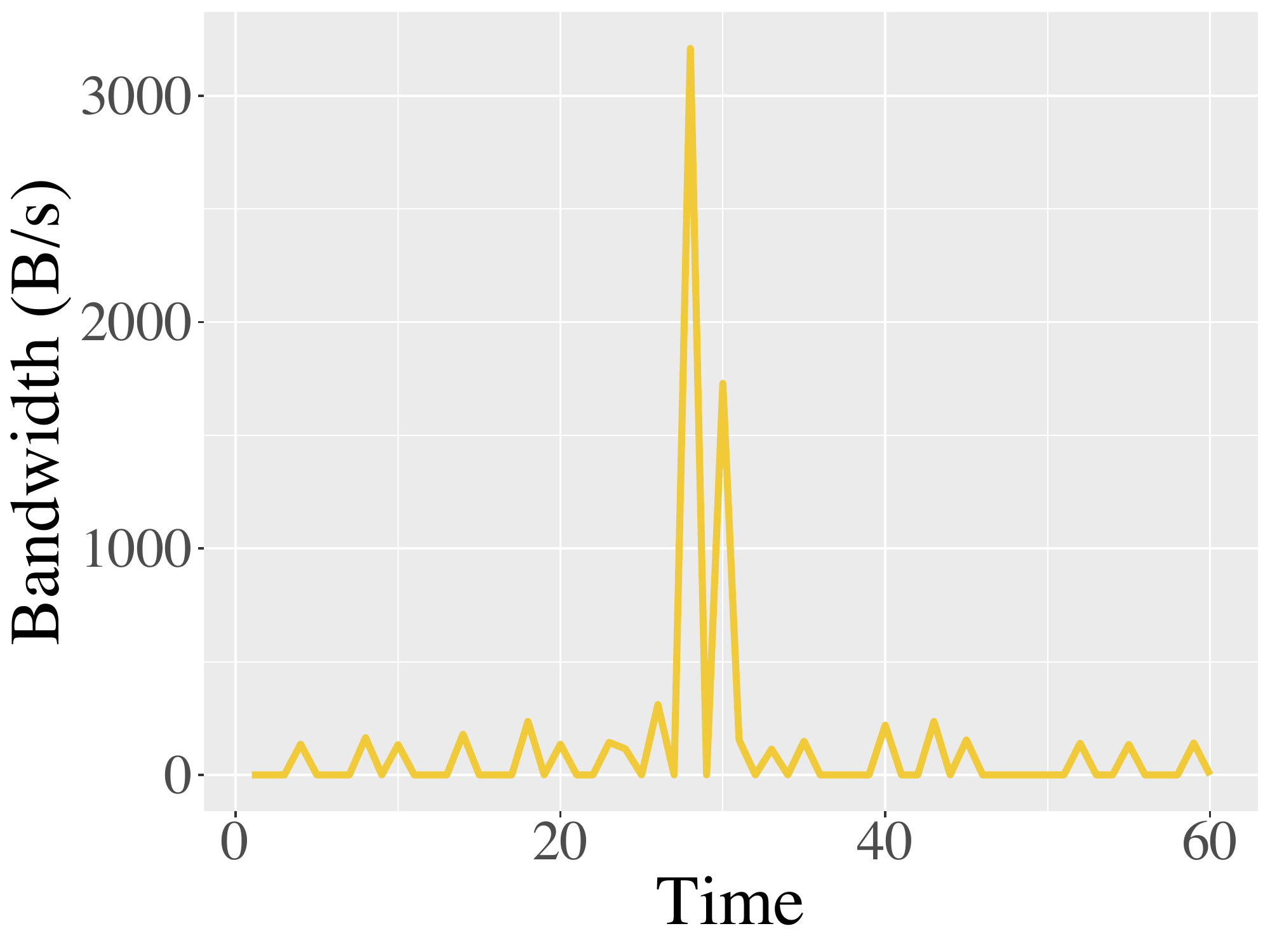}
		\caption{CWcollab.}
		\label{fig:pdf-our-preload}
	\end{subfigure}
	~
	\begin{subfigure}[t]{\evalfigwidth\textwidth}
		\centering
		\includegraphics[width=0.8\linewidth]{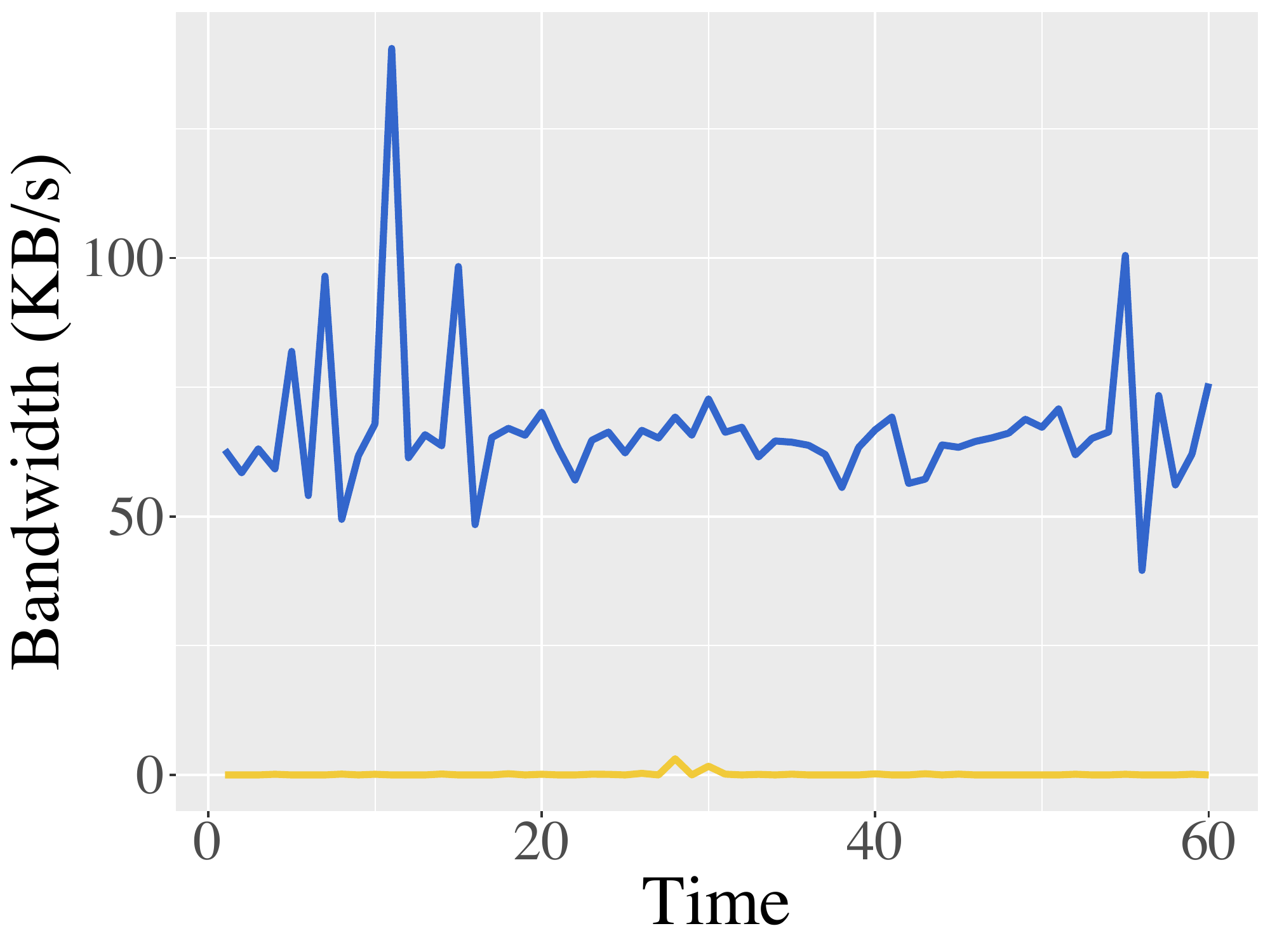}
		\caption{Comparison.}
		\label{fig:pdf-all}
	\end{subfigure}
	\caption{Comparison of network usages when presenting a \textbf{PDF document} with operations such as paging up/down, scrolling and commenting using Google Hangouts and CWcollab.}
	\label{fig:pdf-eval-perf} 
\end{figure*}

\begin{figure*}[thb]
	\centering
	\begin{subfigure}[t]{\evalfigwidth\textwidth}
		\centering
		\includegraphics[width=0.8\linewidth]{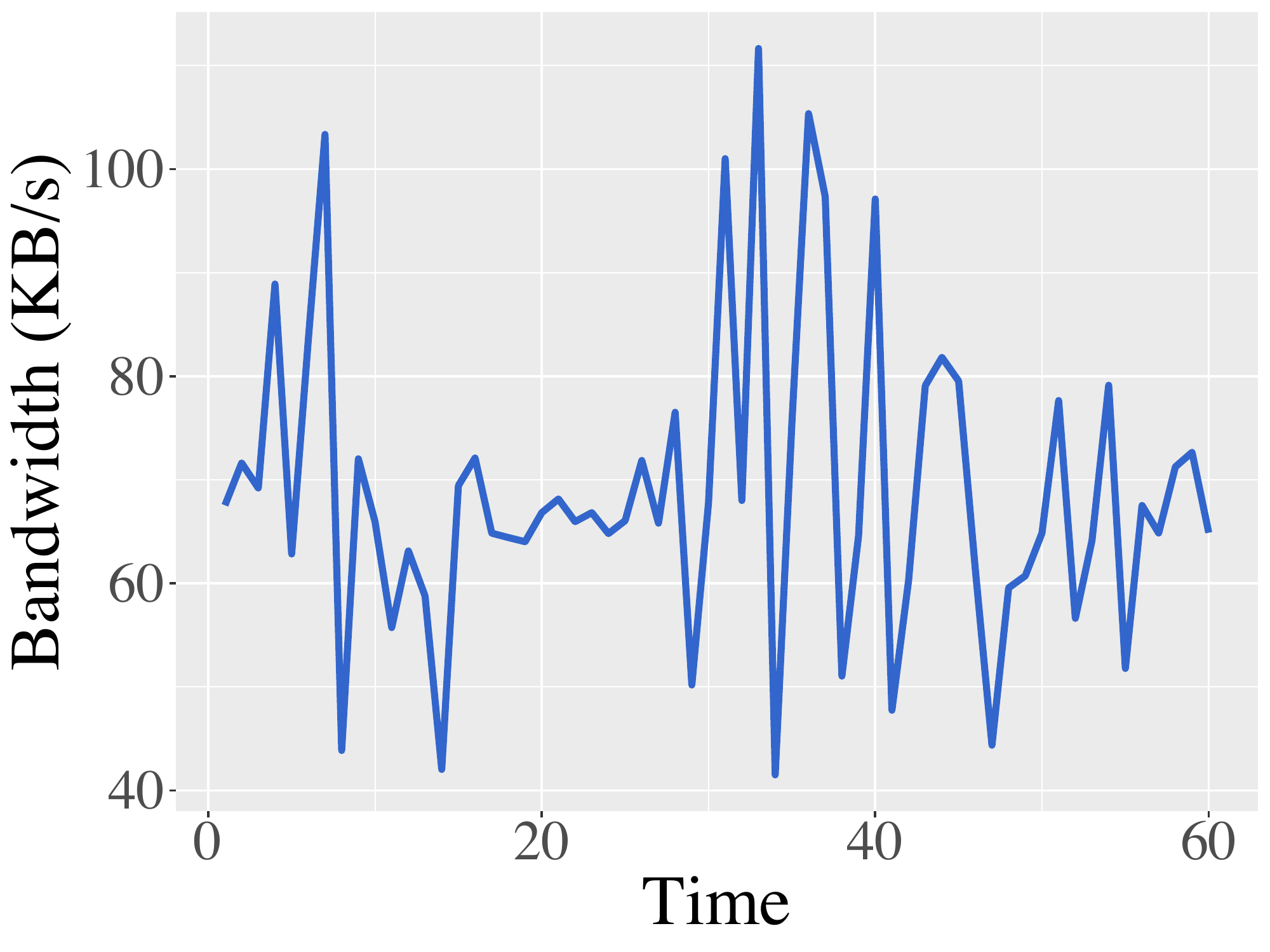}
		\caption{Google Hangouts.}
		\label{fig:image-google-preload}
	\end{subfigure}
	~
	\begin{subfigure}[t]{\evalfigwidth\textwidth}
		\centering
		\includegraphics[width=0.8\linewidth]{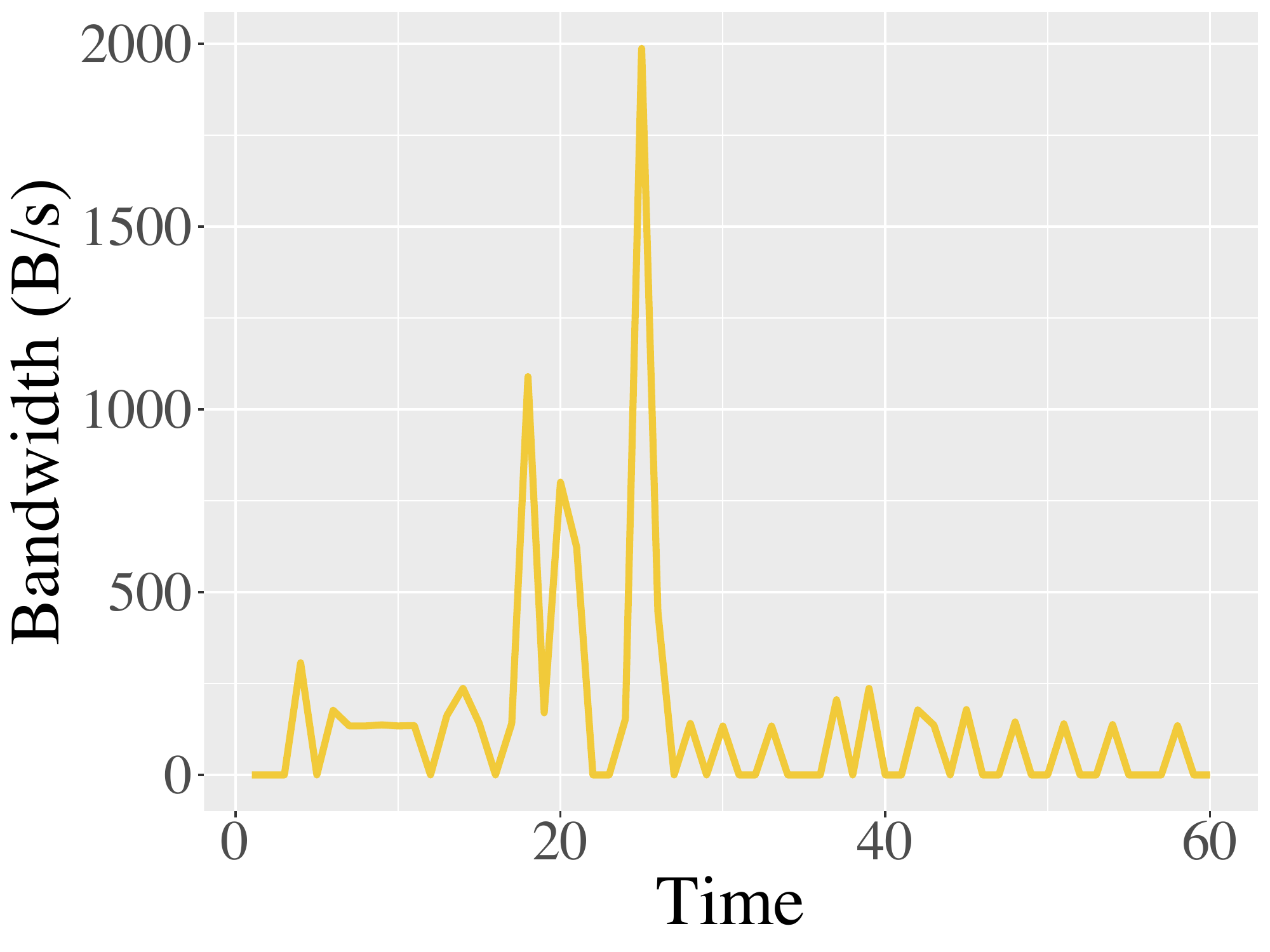}
		\caption{CWcollab.}
		\label{fig:image-our-preload}
	\end{subfigure}
	~
	\begin{subfigure}[t]{\evalfigwidth\textwidth}
		\centering
		\includegraphics[width=0.8\linewidth]{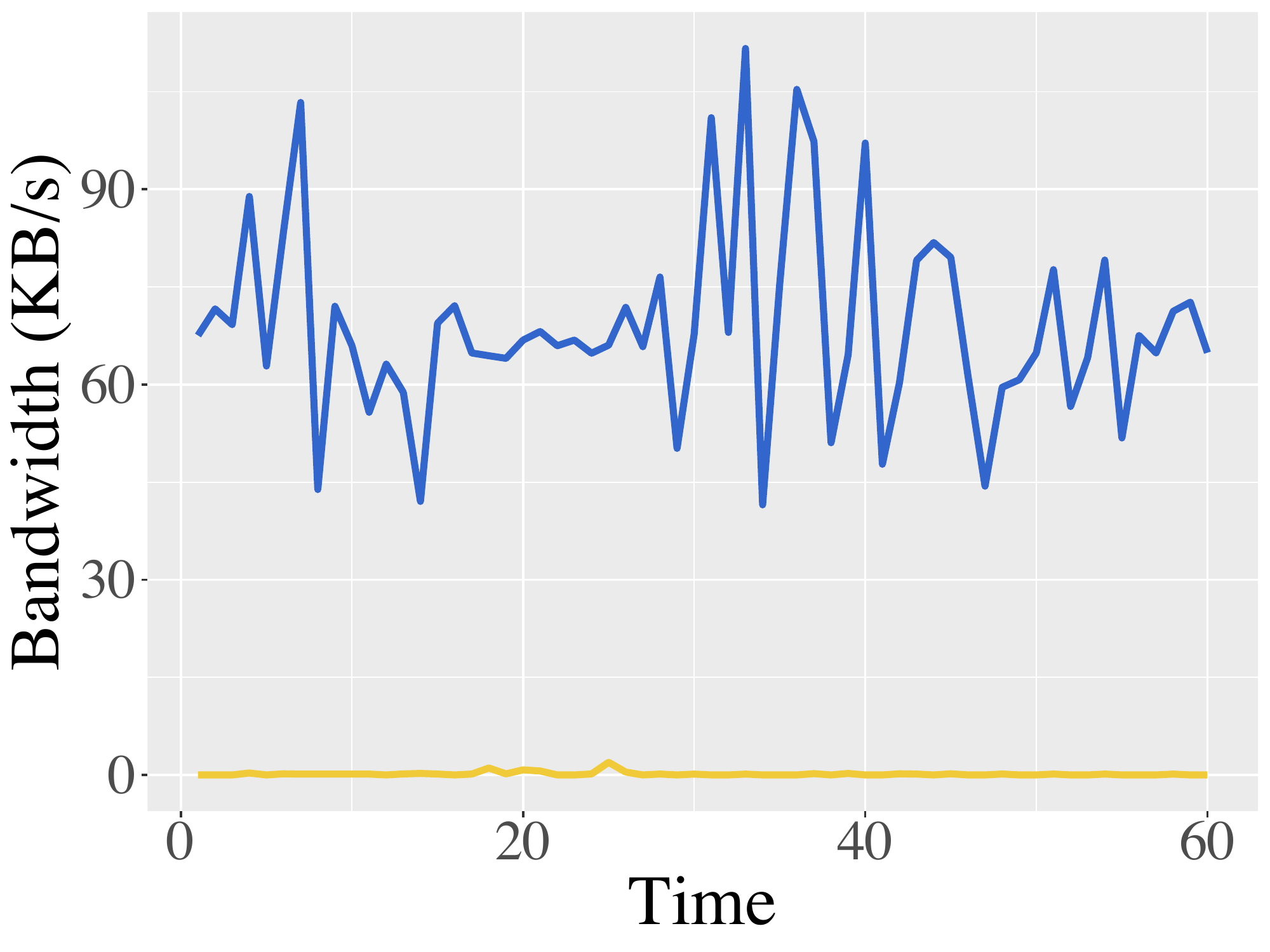}
		\caption{Comparison.}
		\label{fig:image-all}
	\end{subfigure}
	\caption{Comparison of network usages when presenting an \textbf{image} with operations such as zooming in/out, moving, and free drawing using Google Hangouts and CWcollab.}
	\label{fig:image-eval-perf} 
\end{figure*}

\begin{figure*}[thb]
	\centering
	\begin{subfigure}[t]{\evalfigwidth\textwidth}
		\centering
		\includegraphics[width=0.8\linewidth]{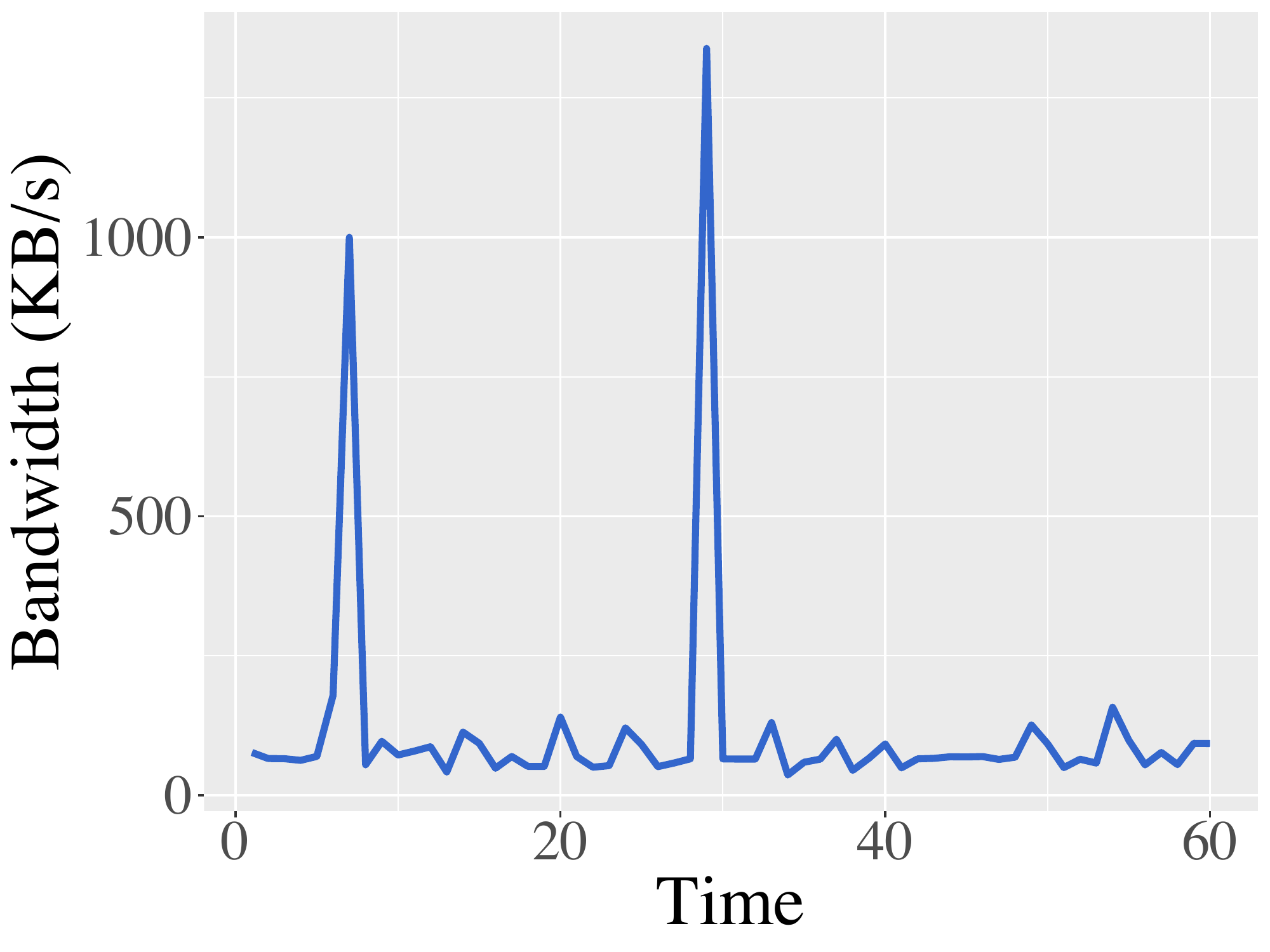}
		\caption{Google Hangouts.}
		\label{fig:webpage-google-afterload}
	\end{subfigure}
	~
	\begin{subfigure}[t]{\evalfigwidth\textwidth}
		\centering
		\includegraphics[width=0.8\linewidth]{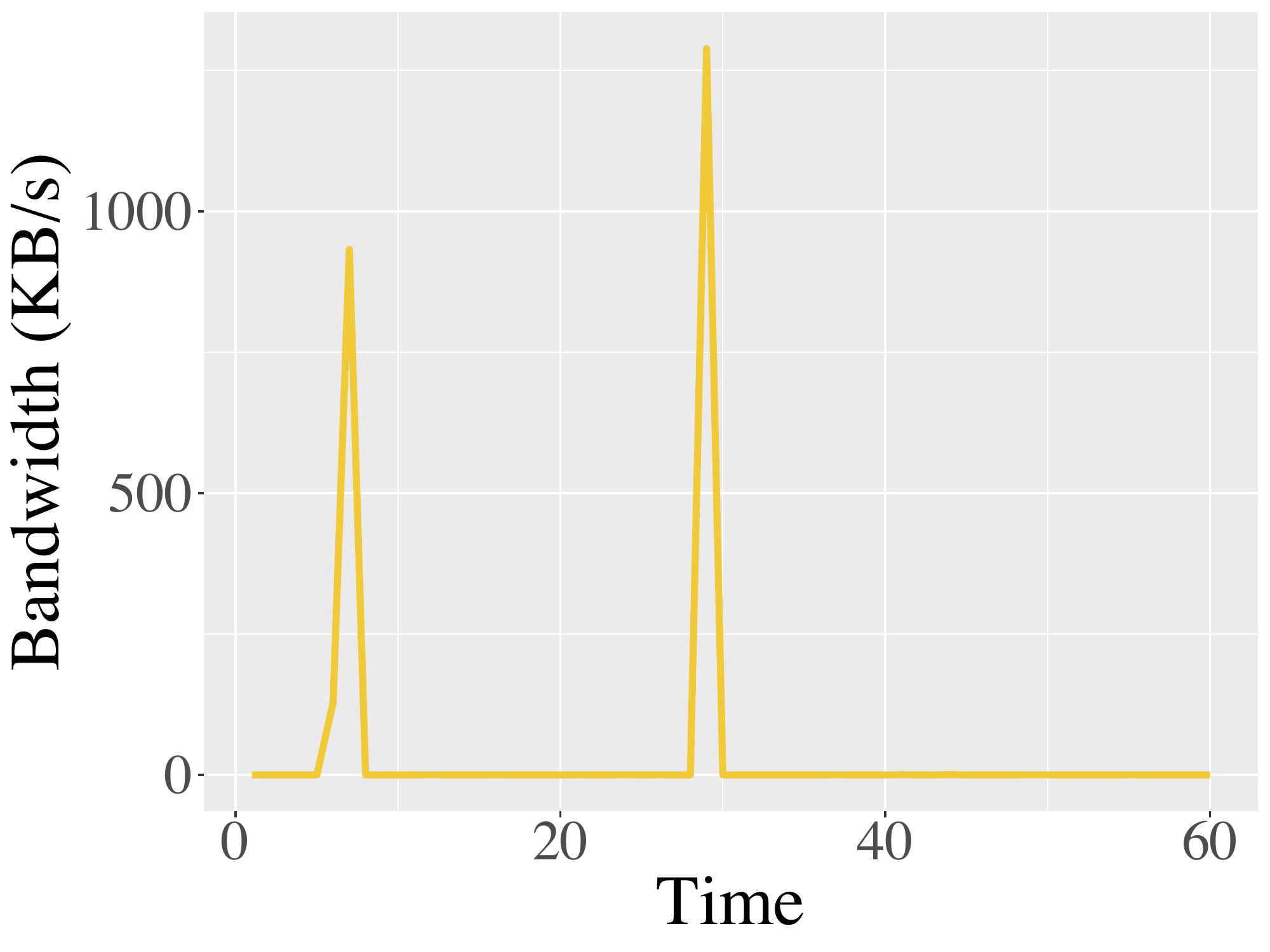}
		\caption{CWcollab.}
		\label{fig:webpage-our-afterload}
	\end{subfigure}
	~
	\begin{subfigure}[t]{\evalfigwidth\textwidth}
		\centering
		\includegraphics[width=0.8\linewidth]{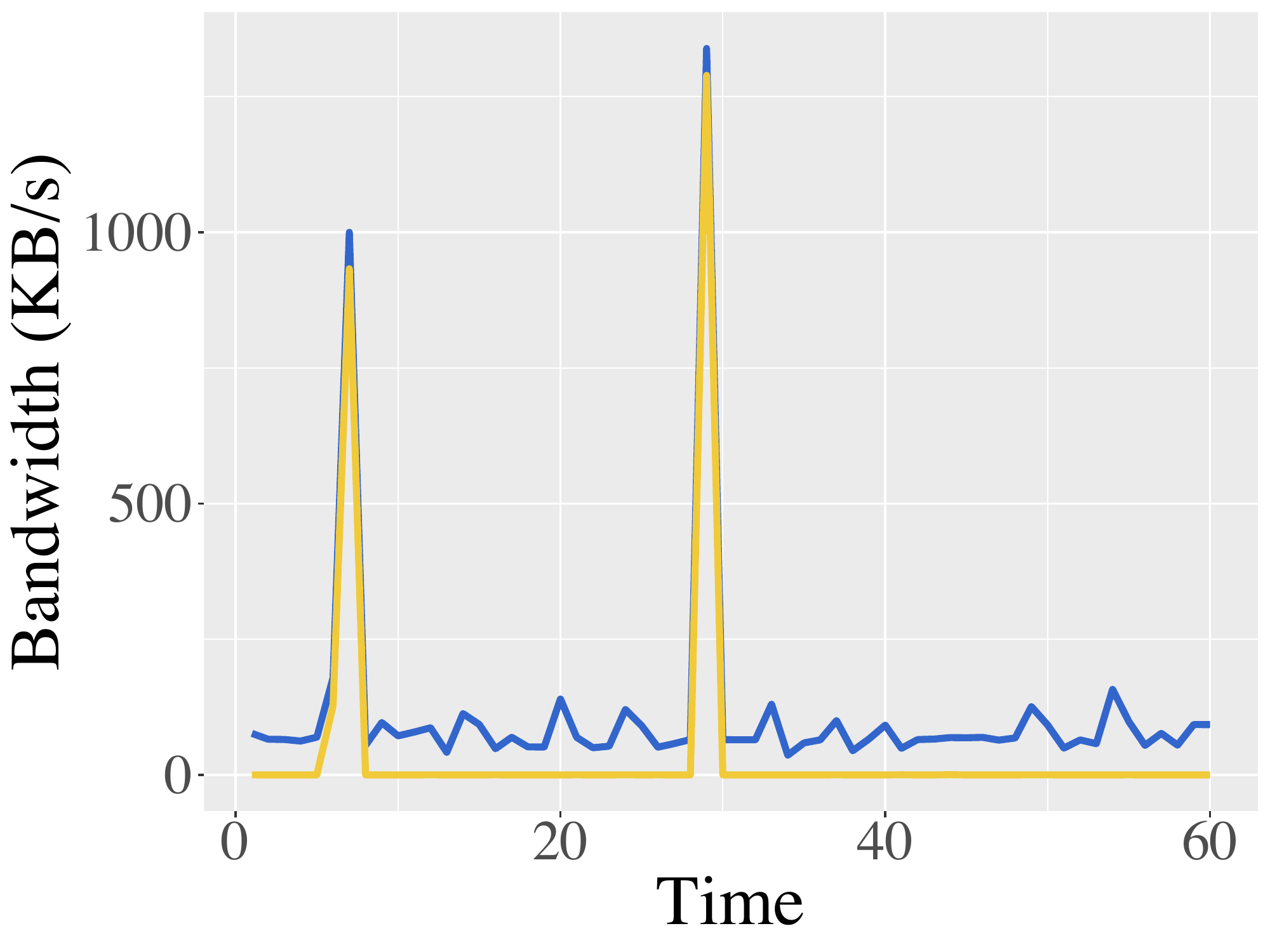}
		\caption{Comparison.}
		\label{fig:webpage-all}
	\end{subfigure}
	\caption{Comparison of network usages when presenting a \textbf{web page} with operations such as clicking links, scrolling, and highlighting using Google Hangouts and CWcollab.}
	\label{fig:webpage-eval-perf} 
\end{figure*}

Currently, screen sharing products such as Zoom and Google Hangouts are widely used to provide basic collaboration support in video conferencing. In this section, we compare Google Hangouts with CWcollab from multiple perspectives. Although we didn't evaluate other video conferencing products in details, we reckon that they share similar mechanisms for screen sharing.

We performed a series of experiments on multimedia, including videos, PDF documents, images, and web pages, to measure the network bandwidth usages in one minute of Google Hangouts and CWcollab, on the presenter side. Instead of transferring video chunks during a presentation, the static materials can be loaded beforehand, and only minimal events information is transferred in a session. We found that CWcollab requires a much lower bandwidth usage, as illustrated in Figure \ref{fig:video-eval-perf}, \ref{fig:pdf-eval-perf}, \ref{fig:image-eval-perf}, and \ref{fig:webpage-eval-perf}. Note that the y-axis units are different in these figures.

Figure \ref{fig:video-eval-perf} shows the network bandwidth consumption of both Google Hangouts and CWcollab for presenting a video. Figure \ref{fig:video-google-preload} shows the Google Hangouts network bandwidth consumption is around 60KB/s - 100KB/s, which is mainly for transferring video frames. However, CWcollab presets the static media-the video-in this experiment, and the network bandwidth consumption is only around 0B/s - 1KB/s in Figure \ref{fig:video-our-preload}. The bandwidth consumption comparison is summarized in Figure \ref{fig:video-all}. Similarly, Figure \ref{fig:pdf-eval-perf} shows Google Hangouts utilizes around 60KB/s bandwidth while CWcollab only needs a maximum 3KB/s for presenting a PDF document; Figure \ref{fig:image-eval-perf} shows Google Hangouts utilizes around 70 KB/s and CWcollab only needs a maximum 2KB/s for presenting an image; Figure \ref{fig:webpage-eval-perf} shows Google Hangouts needs around 100KB/s and CWcollab needs almost 0B/s most of the time (the two peaks are for loading web pages). To summarize Figure \ref{fig:video-eval-perf}-\ref{fig:webpage-eval-perf}, we can see that in the measurement of one minute, most events only require hundreds of bytes per second. 
The most expensive action is the \textit{free drawing} event whose bandwidth usages can be as high as around 3KB/s. However, the usages are still much lower than those of Google Hangouts.

We also summed up the bytes transmitted in this one minute and showed the results in Table~\ref{tab:bytes}. CWcollab transmits very few bytes in a presentation, especially noticing that in the collaboration session except web pages, in one minute, the total bytes transmitted in CWcollab is less than \textbf{10 KB}, because CWcollab leverages an object-prioritized approach to capture media events and represent them in simple messages. As the web pages have to be loaded dynamically, even though loading pages costs a large amount of network bandwidth, CWcollab still requires much lower bandwidth than Google Hangouts.

\begin{table}[!htb]
	\caption{Total bytes transmitted in one minute of four types of media.}
	\label{tab:bytes}
	\centering
	\begin{tabular}{|l|c|c|c|c|}
		\hline
		& \textbf{Video} & \textbf{PDF} & \textbf{Image} & \textbf{Web page} \\ \hline
		Google Hangouts   & 3.80MB  & 4.01MB & 4.04MB  & 6.58MB  \\ \hline
		CWcollab          & 6.73KB  & 7.76KB & 8.80KB  & 2.30MB        \\ \hline
	\end{tabular}
\end{table}

Besides bandwidth usages, there are some other differences between screen sharing tools and CWcollab. Because the mechanism of screen sharing tools is to capture and broadcast the display, it cares nothing about what application the user is synchronizing. As a result, it is straightforward for these tools to support multimedia in web browsers and desktop environment. By contrast, our platform is web-based. To support another type of media, we need to instrument necessary control to achieve the collaboration. Additionally, CWcollab can be a supplement to current screen sharing products, considering that it provides an efficient platform to support message-based collaboration on multimedia.

%% file: text/conclusion.tex
\section{Conclusion}\label{sec.conclusion}

In this paper, we proposed a context-aware web-based collaborative multimedia  system-CWcollab. It supports collaboration on multimedia and uses simple messages to represent media controls. We demonstrated the design methodologies of a distributed collaboration framework and discussed the implementation of architectural components. We compared CWcollab with screen sharing tools such as Google Hangouts. Our evaluation results showed that our tool has a lower bandwidth usage than that of Google Hangouts. We posit that this work could serve as a uniform platform for efficient general-purpose multimedia collaboration.

%% file: main.bbl
\begin{thebibliography}{10}
\providecommand{\url}[1]{#1}
\csname url@samestyle\endcsname
\providecommand{\newblock}{\relax}
\providecommand{\bibinfo}[2]{#2}
\providecommand{\BIBentrySTDinterwordspacing}{\spaceskip=0pt\relax}
\providecommand{\BIBentryALTinterwordstretchfactor}{4}
\providecommand{\BIBentryALTinterwordspacing}{\spaceskip=\fontdimen2\font plus
\BIBentryALTinterwordstretchfactor\fontdimen3\font minus
  \fontdimen4\font\relax}
\providecommand{\BIBforeignlanguage}[2]{{%
\expandafter\ifx\csname l@#1\endcsname\relax
\typeout{** WARNING: IEEEtran.bst: No hyphenation pattern has been}%
\typeout{** loaded for the language `#1'. Using the pattern for}%
\typeout{** the default language instead.}%
\else
\language=\csname l@#1\endcsname
\fi
#2}}
\providecommand{\BIBdecl}{\relax}
\BIBdecl

\bibitem{greenberg1990group}
S.~Greenberg and R.~Bohnet, ``Group sketch: a multi-user sketchpad for
  geographically-distributed small groups,'' 1990.

\bibitem{tang1991videowhiteboard}
J.~C. Tang and S.~Minneman, ``Videowhiteboard: video shadows to support remote
  collaboration,'' in \emph{Proceedings of the SIGCHI Conference on Human
  Factors in Computing Systems}.\hskip 1em plus 0.5em minus 0.4em\relax ACM,
  1991, pp. 315--322.

\bibitem{elrod1992liveboard}
S.~Elrod, R.~Bruce, R.~Gold, D.~Goldberg, F.~Halasz, W.~Janssen, D.~Lee,
  K.~McCall, E.~Pedersen, K.~Pier \emph{et~al.}, ``Liveboard: a large
  interactive display supporting group meetings, presentations, and remote
  collaboration,'' in \emph{Proceedings of the SIGCHI conference on Human
  factors in computing systems}.\hskip 1em plus 0.5em minus 0.4em\relax ACM,
  1992, pp. 599--607.

\bibitem{goldman2011real}
M.~Goldman, G.~Little, and R.~C. Miller, ``Real-time collaborative coding in a
  web ide,'' in \emph{Proceedings of the 24th annual ACM symposium on User
  interface software and technology}.\hskip 1em plus 0.5em minus 0.4em\relax
  ACM, 2011, pp. 155--164.

\bibitem{yoon2016richreview++}
D.~Yoon, N.~Chen, B.~Randles, A.~Cheatle, C.~E. L{\"o}ckenhoff, S.~J. Jackson,
  A.~Sellen, and F.~Guimbreti{\`e}re, ``Richreview++: Deployment of a
  collaborative multi-modal annotation system for instructor feedback and peer
  discussion,'' in \emph{Proceedings of the 19th ACM Conference on
  Computer-Supported Cooperative Work \& Social Computing}.\hskip 1em plus
  0.5em minus 0.4em\relax ACM, 2016, pp. 195--205.

\bibitem{wenzel2016full}
M.~Wenzel and C.~Meinel, ``Full-body webrtc video conferencing in a web-based
  real-time collaboration system,'' in \emph{2016 IEEE 20th International
  Conference on Computer Supported Cooperative Work in Design (CSCWD)}.\hskip
  1em plus 0.5em minus 0.4em\relax IEEE, 2016, pp. 334--339.

\bibitem{kim2015interactive}
K.~Kim, W.~Ha, O.~Choi, H.~Yeh, J.-H. Kim, M.~Hong, and T.~Shon, ``An
  interactive pervasive whiteboard based on mvc architecture for ubiquitous
  collaboration,'' \emph{Multimedia Tools and Applications}, vol.~74, no.~5,
  pp. 1557--1576, 2015.

\bibitem{booth2002mighty}
K.~S. Booth, B.~D. Fisher, C.~J.~R. Lin, and R.~Argue, ``The mighty mouse
  multi-screen collaboration tool,'' in \emph{Proceedings of the 15th annual
  ACM symposium on User interface software and technology}.\hskip 1em plus
  0.5em minus 0.4em\relax ACM, 2002, pp. 209--212.

\bibitem{bentley1997designing}
R.~Bentley and W.~Appelt, ``Designing a system for cooperative work on the
  world-wide web: Experiences with the bscw system,'' in \emph{Proceedings of
  the Thirtieth Hawaii International Conference on System Sciences, 1997},
  vol.~4.\hskip 1em plus 0.5em minus 0.4em\relax IEEE, 1997, pp. 297--306.

\bibitem{bentley1997world}
R.~Bentley, T.~Horstmann, and J.~Trevor, ``The world wide web as enabling
  technology for cscw: The case of bscw,'' \emph{Computer Supported Cooperative
  Work (CSCW)}, vol.~6, no. 2-3, pp. 111--134, 1997.

\bibitem{mogan2010impact}
S.~Mogan and W.~Wang, ``The impact of web 2.0 developments on real-time
  groupware,'' in \emph{2010 IEEE Second International Conference on Social
  Computing (SocialCom)}.\hskip 1em plus 0.5em minus 0.4em\relax IEEE, 2010,
  pp. 534--539.

\bibitem{wang2009cooperative}
W.~Wang, K.~Finch, J.~Rubart, and J.~M. Haake, ``A cooperative hypermedia
  approach to flexible process support for managing distributed projects,''
  \emph{International Journal of Cooperative Information Systems}, vol.~18, no.
  03n04, pp. 481--512, 2009.

\bibitem{schmid2014real}
O.~Schmid, A.~L. Masson, and B.~Hirsbrunner, ``Real-time collaboration through
  web applications: an introduction to the toolkit for web-based interactive
  collaborative environments (twice),'' \emph{Personal and Ubiquitous
  Computing}, vol.~18, no.~5, pp. 1201--1211, 2014.

\bibitem{fetter2014lightweight}
M.~Fetter, R.~Strobel, and T.~Gross, ``Lightweight support for collaborative
  web browsing through spreadvector,'' in \emph{Proceedings of the extended
  abstracts of the 32nd annual ACM conference on Human factors in computing
  systems}.\hskip 1em plus 0.5em minus 0.4em\relax ACM, 2014, pp. 1339--1344.

\bibitem{binda2018phamilyhealth}
J.~Binda, E.~Georgieva, Y.~Yang, F.~Gui, J.~Beck, and J.~M. Carroll,
  ``Phamilyhealth: A photo sharing system for intergenerational family
  collaboration on health,'' in \emph{Companion of the 2018 ACM Conference on
  Computer Supported Cooperative Work and Social Computing}.\hskip 1em plus
  0.5em minus 0.4em\relax ACM, 2018, pp. 337--340.

\bibitem{lee2018exploring}
S.~W. Lee, R.~Krosnick, S.~Y. Park, B.~Keelean, S.~Vaidya, S.~D. O'Keefe, and
  W.~S. Lasecki, ``Exploring real-time collaboration in crowd-powered systems
  through a ui design tool,'' \emph{Proceedings of the ACM on Human-Computer
  Interaction}, vol.~2, no. CSCW, p. 104, 2018.

\bibitem{kunz2010collaboard}
A.~Kunz, T.~Nescher, and M.~Kuchler, ``Collaboard: a novel interactive
  electronic whiteboard for remote collaboration with people on content,'' in
  \emph{2010 International Conference on Cyberworlds (CW)}.\hskip 1em plus
  0.5em minus 0.4em\relax IEEE, 2010, pp. 430--437.

\end{thebibliography}
